\documentclass[10pt]{article}
\makeatletter
\def\u{u}
\def\F{F}
\def\d{\,{\rm d}}
\def\e{{\rm e}}
\def\w{\wedge}

\newcommand{\ph}[1]{{\tilde\phi}_1^{#1}}
\newcommand\phys[3]{{\it Phys. Rev.} {\rm D} {\bf {#1}}, 
{#2} ({#3})}
\newcommand{\Fp}[1]{F^{#1}_{\varphi z}}
\newcommand{\der}[1]{{\frac{\d {#1}}{\d r}}}
\newcommand{\derm}[1]{{\frac{\d {#1}}{\d m}}}
\newcommand{\nder}[3]{{\frac{\d^{#1} {#2}}{\d {#3}^{#1}}}}
\newcommand{\pder}[2]{\frac{\partial {#1}}{\partial {#2}}}
\newcommand\eref[1]{(\ref{#1})}
\newcommand{\T}[1]{{\Theta}^{{\hat #1}}}
\def\eqnarray{%
   \stepcounter{equation}%
   \def\@currentlabel{\p@equation\theequation}%
   \global\@eqnswtrue
   \m@th
   \global\@eqcnt\z@
   \tabskip\@centering
   \let\\\@eqncr
   $$\everycr{}\halign to\displaywidth\bgroup
       \hskip\@centering$\displaystyle\tabskip\z@skip{##}$\@eqnsel
      &\global\@eqcnt\@ne \hfil${{}##{}}$\hfil
      &\global\@eqcnt\tw@
         $\displaystyle{##}$\hfil\tabskip\@centering
      &\global\@eqcnt\thr@@ \hb@xt@\z@\bgroup\hss##\egroup
         \tabskip\z@skip
      \cr
}
\newenvironment{subequations}{%
  \refstepcounter{equation}%
  \mathchardef\c@mainequation\c@equation
  \protected@edef\themainequation{\theequation}%
  \let\theequation\thesubequation
  \global\c@equation\z@
  }{%
  \global\c@equation\c@mainequation
  \global\@ignoretrue
  }
\newcommand{\thesubequation}{\themainequation\alph{equation}}

\newcommand*{\@lab@subeqnarray}[2]{#1{#2}\eqnarray}
\@addtoreset{equation}{section}
\def\theequation{\@arabic\c@section.\@arabic\c@equation}
\makeatother


\begin{document}

\begin{center}
{\bf \Large A cylindrically symmetric solution in\\[2mm] 
Einstein-Maxwell-dilaton gravity\footnote[4]
{Suggested running head: 
Cylindrically symmetric solution in EMD gravity}
}\\[3mm]
P. Klep\'a\v{c}
\footnote[1]
{Institute of Theoretical Physics and 
Astrophysics, Faculty of Science,\\ Masaryk 
University, Kotl\'{a}\v{r}sk\'{a} 2, 611 37  Brno, 
Czech Republic}
\footnote[2]
{E-mail: {\tt klepac@physics.muni.cz}} 
and J. Horsk\'y$^{\ 1}$
\footnote[3]{E-mail: 
{\tt horsky@physics.muni.cz}}\\[2mm]
\end{center} 
{\small We consider the existence of 
Einstein-Maxwell-dilaton plus fluid system 
for the case of stationary 
cylindrically symmetric spacetimes. An exact 
inhomogeneous $ \varepsilon$-order solution 
is found, where the parameter $\varepsilon$ parametrizes 
the non-minimally coupled electromagnetic field. 
Some its physical attributes are investigated 
and a connection with already known G\"odel-type 
solution is given. 
It is shown that the found solution also survives 
in the string-inspired charged gravity framework. 
We find that a magnetic field has positive 
influence on the chronology violation unlike 
the dilaton influence.\\

\noindent KEY WORDS: exact solutions, charged perfect 
fluid, scalar field}

\newpage

\section{\label{sec1}Introduction}

Einstein's theory of relativity is, in general, an excellent 
approximation of gravitational phenomena which appear at 
low energies. Nevertheless, if one goes to energies at 
the Planck scale, then one is faced with the neccessity of 
introducing quantum corrections. Nowadays, superstring theory 
is believed to unify succesfully all of the known
fundamental interactions observed in nature. Moreover the 
original Einstein's theory naturally emerges if one 
ignores all higher-order stringy corrections. 

In the last decade the string cosmology has become 
an attractive subject of interest. As one goes to 
the low energies, string cosmology is actually the 
classical cosmology with general relativity enriched 
by addition of massless scalar fields. An incorporation 
of these fields is perhaps promissible way how 
to resolve long standing problems in cosmology. 

A number of solutions to the so called 
Einstein-Maxwell-dilaton gravity, i.e. 
relativistic gravity theory containing 
non-minimally coupled dilaton and electromagnetic 
fields, has been derived by various techniques \cite{EMD}. 
Barrow and D\c{a}browski 
\cite{Barrow} obtained stringy G\"odel-type solution 
without closed timelike curves (CTC's) 
by considering the one-loop corrected 
superstring effective action. Kanti and Vayonakis 
\cite{Kanti} have extended analysis of \cite{Barrow} 
on case with an electromagnetic field, and their 
found that the chronology is violated. Also results 
of others authors show that presence of the electromagnetic 
field may cause the chronology violation (especially 
purely magnetic filed parallel to the rotation axis) 
while a scalar field may again restore the chronology 
\cite{{Reb},{klep}}.
   
In this work some results on the stationary cylindrically 
symmetric spacetimes in Einstein-Maxwell-dilaton (EMD) theory 
of gravity are presented. The reason for studying this 
class of the spacetimes is twofold. First, searching for 
EMD solutions is important in itself. 
Second, in the 
classical relativity theory the cylindrically symmetric 
spacetimes are known to violate some of the chronology 
conditions \cite{Hawking}. Therefore it is natural to 
address the question of chronology violation in the EMD
spacetimes. The paper extends 
the results of \cite{{Barrow},{Kanti}} to the inhomogeneous 
case where, in general, only three isometries are present. 

The paper is organized as follows. After 
some preliminaries in section \ref{sec2} 
we derive in section \ref{sec3} the exact solution 
for the lowest order in parameter $\varepsilon$, 
which parametrizes the electromagnetic field.
In section 
\ref{sec4} there are the results of the previous 
one generalized on the 
$\varepsilon$-order corrections in the framework of the 
EMD theory. The case with more scalar 
fields is studied in section \ref{sec5} provided 
that in zero-order they depend solely on 
the longitudinal direction. In section \ref{sec8} 
it is shown that the found solution in fact 
still applies even if string-inspired charged 
gravity is taken under consideration. 
Finally 
section \ref{sec6} briefly summarizes the basic properties 
of this solution.

\section{\label{sec2}Preliminaries}

We search for cylindrically symmetric 
stationary spacetimes. Then there exist 
local coordinate systems $(x^0,x^1,x^2,x^3)=
(t,\varphi,z,r)$ adapted to Killing fields 
$\partial_t,\ \partial_\varphi,\ \partial_z$, where 
the hypersurfaces $\varphi=0$ and $\varphi=2\pi$ are to be 
identified and $\partial_t$ is everywhere a nonvanishing 
timelike field.

Furthermore we choose a local coframe fields 
$\T\mu$ defined by (Greek indices 
run from 0 to 3)
\begin{equation}\label{basis}
\begin{array}{ll}
\T 0 = \e^\alpha(\d t+f\d\varphi)\ ,  \qquad & 
\T 1 = l\d\varphi\ , \\
\T 2 = \d z\ , \qquad \qquad &  
\T 3 = \e^\delta\d r\ ,
\end{array}
\end{equation}
with $f,\ l,\ \alpha,\ \delta$ 
being functions of $r$ only.

Let the metric tensor field be in the basis 
\eref{basis} written as 
$g=\eta_{\mu\nu}\T\mu\otimes\T\nu$, where 
$(\eta_{\mu\nu})={\rm diag}(\ 1,\ -1,\ -1,\ -1)$ 
is the Minkowski matrix. 

The defining equations 
\eref{basis} show that in this paper 
the $g_{zz}$ metric field component is constant, 
in contrast to \cite{{klep},{klep2}}, where the case 
with constant $g_{tt}$ component was studied.

Let the spacetime is filled with the 
charged perfect fluid and massless scalar field $\phi$.
The perfect fluid is characterized by its pressure 
$p$ and energy density $\mu$, from symmetry considerations 
both these quantities depending only on the radial 
coordinate. 
On the other hand $\phi$ may depend 
also on the longitudinal direction $z$.
The electromagnetic field is non-minimally coupled 
to $\phi$ with parameter $\varepsilon$. 
Here we would like to point out that the spacetime 
with basis fields \eref{basis} is referred to cylindrically 
symmetric althought the dilaton generally depends 
also on the longitudinal direction.

The $\varepsilon$-order 
EMD action that we 
will deal with is given by 
\begin{equation}
S_{\rm eff}[g,A,\phi,\mu]=\intop_{\cal M}\left[*R
+16\pi*\mu-\d\phi\wedge *\d\phi
+2\varepsilon\ \e^{\phi}\ \F\wedge*\F \right]\ , 
\label{stringaction} 
\end{equation}
where $\varepsilon$ is a real parameter, 
$R$ is the Ricci scalar of the 
metric tensor, $\F$ is the electromagnetic 
field 2-form. The scalar field $\phi$ 
will be henceforth called dilaton. 

We claim to obtain a EMD solution which is 
of the first order in the parameter $\varepsilon$. 
Of course at the first place it means 
one should have a zero-order solution in $ \varepsilon$ 
that solves equations of motion for the action 
\eref{stringaction} if we let $ \varepsilon$ to be zero, 
or in other words, a purely classical solution of the 
Einstein equations coupled with a scalar field 
and perfect fluid. Then the $ \varepsilon$-order 
solution we are looking for is naturally viewed as 
electromagnetic 
first order correction of the classical solution, and it 
has to coincide with the latter when $ \varepsilon$ goes to 
zero.

As for the dilaton $ \phi$, it may be written as the 
sum of zero-order solution $\phi^{(0)}$ and a $\varepsilon$-order 
correction like
\begin{equation}\label{dilcor}
\phi=\phi^{(0)}+\varepsilon\phi^{(1)}\ .
\end{equation}
If only terms linear in $ \varepsilon$ are considered, one is 
forced to keep terms of $ \varepsilon$-order in the corresponding 
equations of motion. Particularly, the coupling fuction 
standing at $F\w*F$, being already of ${\cal O}(\varepsilon
)$ becomes in our approximation equal to 
$2 \varepsilon \e^{\phi^{(0)}}$.

An usual progress is to introduce a fluid comoving system, 
in which the fluid particles motion is uniquely determined 
by the velocity (co)vector field $u$, $u=\T 0$. 
Let us very briefly mention basic properties of the geometry 
of fluid particles worldlines congruences. An acceleration 
1-form $\dot u$ 
is given as ${\dot u}=- \d \alpha$. Since the problem is 
stationary, both expansion and shear tensor are vanishing.  
A vorticity covector is given by
\begin{equation}\label{vorticity}
 \omega=\frac12*(u\w\d u)=\frac12\der f l^{-1}
\e^{\alpha-\delta}\d z\ .
\end{equation}
The last two statements show that the fermionic fluid rotates as a
rigid body.

\section{\label{sec3}Zeroth-order solution}

In this paragraph we derive a solution of zero-order in 
$ \varepsilon$. In this case the action \eref{stringaction} 
becomes Einstein-dilaton plus fluid system. The Einstein field 
equations written in the tetrad representation 
\eref{basis} then read \begin{equation}\label{Einsteinframe}
-\frac12\ {\eta}_{\alpha\beta\gamma}\w
{\Omega}^{{\hat\beta}{\hat\gamma}}=
8\pi*i_{\hat\alpha}{T}\ ,
\end{equation}
where $i_{\hat\alpha}\equiv i_{{\bf e}_{\hat\alpha}}$ is 
the interior product (${\bf e}_{\hat\alpha}$ is dual basis to 
\eref{basis}, $\T \alpha 
({\bf e}_{\hat\beta})=\delta^\alpha_\beta$), 
$\Omega$ is curvature 2-form on 
$T{\cal M}$, ${\Omega}^{\hat\alpha}_{\hat\beta}=\frac 12
R^{\hat\alpha}_{~{\hat\beta}{\hat\gamma}\hat\delta}
\T\gamma\w\T\delta$, and 1-form $\eta^{\alpha \beta\gamma}$ 
is defined by \cite{Straumann}
\[{\eta}^{\alpha\beta\gamma}=*
(\T\alpha\w\T\beta\w\T\gamma)\ .\]
Finally $T$ is the total stress-energy tensor of the 
perfect fluid and the masseless scalar field, 
\begin{equation}
\begin{array}{rcl}
8\pi T&=&8\pi\left[(\mu+p)\u\otimes\u
-p\ g\right]\\
&+&\frac12\left[\d \phi\otimes\d\phi-
\frac12\ g(\d\phi,\d\phi)\ g\right]\ .
\label{stress-energy}
\end{array}
\end{equation}

Explicit form of the Einstein equations 
takes the form
\begin{subequations}\label{Einsteineqscl}
\begin{eqnarray} 
&\displaystyle{\frac{\d}{\d r}\left[\frac {\e^{2\alpha}}M
\der f\right]=0\ ,}
  \label{omegacl}\\[2mm]
&\displaystyle{\!\frac 2M\frac{\d}{\d r}
\left[\frac 1M\der \alpha \right]
  \!=-\!\frac 1{M^2}\!\left(\pder{\phi}{r}\right)^{\!\!2}\!\!\ ,}
  \label{ff-rrcl} \\[2mm]
&\displaystyle{\frac 1M\frac{\d}{\d r}\left[
\frac {\e^{-2\alpha }}M \frac{\d }{\d r}
\left(l^2\e^{2\alpha }\right)\right]=
  32\pi p\, \e^{2\alpha}\ ,}
  \label{zz+rrcl}\\[2mm]
&\displaystyle
{l^2\pder\phi r\pder\phi z= 0\ ,}
  \label{zrcl}\\[2mm]
&\displaystyle{\frac{\e^{-2\alpha}}M{\frac{\d}{\d  r}
\!\left[\frac1M
  \frac{\d l^2}{\d r}\right]=}
  {16\pi(p-\mu)
  +\frac 1{M^2}\left(\der f\right)^{\!\!2}}\!
-\left(\pder {\phi}{z}\right)^{\!\!2}\!\ ,}
  \label{tt-rrcl} \\
&\displaystyle{2\e^{-2\alpha}\der\alpha 
\frac{\d \, l^2}{\d r}=
32\pi M^2 p- 
  \left(\der f\right)^{\!\!2}+\ l^2
\left(\pder {\phi}r\right)^{\!\!2}\e^{-2\alpha}
  -\left(\pder {\phi}z\right)^{\!\!2}M^2\ ,}
  \label{intBianchicl}
\end{eqnarray}
\end{subequations}

Let us introduce the functions 
$m$ and $M$ by the formulae 
\[M=l\e^{\delta-\alpha}\ ,\qquad m=\int M\d r\ .\]

The scalar field equation of motion is the 
massless Klein-Gordon equation
\begin{equation}\label{Klein-Gordon}
*\ \d*\d\ \phi=\frac 1M\frac\partial{\partial r}\left(\frac
{l^2}{M}\pder\phi r\right)+\frac{\partial^2 \phi}
{\partial z^2}=0\, .
\end{equation}
The Bianchi identity, provided that the dilatonic equation of motion 
\eref{Klein-Gordon} is satisfied, becomes 
\begin{equation}
u\w*\left[\mu\d u+\d (pu)\right]=0\ .\label{Bianchicl}
\end{equation}
Because of the independence of $r$ and $z$ coordinates 
in \eref{Einsteineqscl} one can carry out the 
separation of variables in 
\eref{Klein-Gordon} to obtain $\phi$ in terms 
of the metric functions,
\begin{equation}\label{dilaton}
\phi = \phi_0+\phi_1z+\phi_2\int\frac M{ l^2}\ \d r\,,
\quad \phi _1\phi _2=0{}\ ,
\end{equation}
with constants $\phi_0,\ \phi_1$ and $\phi_2$.

Thus one has reduced the problem to 
solving five equations \eref{omegacl}-\eref{intBianchicl} 
minus \eref{zrcl} 
for six unknowns: $ \alpha,\ f,\ l,\ \delta$ 
and physical quantities of pressure $p$ and 
mass (energy) density $\mu$.

Inserting of \eref{omegacl}, \eref{ff-rrcl} and 
\eref{zz+rrcl} into \eref{intBianchicl} yields system 
of two second-order 
equations for $ \alpha$ and $l^2$ that reads 
\begin{subequations}\label{nonlinclass}
\begin{eqnarray}  
&\displaystyle{2l^4\nder2{\alpha }m=-\phi^{~2}_2\ ,}
\label{nonlinalphacl}\\ 
&\displaystyle{\nder2{l^2}m=\phi^{~2}_1\e^{2\alpha }
+4\Omega^2\e^{-2\alpha}\ ,}
\label{nonlinlcl}
 \end{eqnarray}
\end{subequations}

The authors have been able to find a solution 
to \eref{nonlinclass} if $\phi_2=0$, which from 
\eref{dilcor} and \eref{dilaton} immediately implies       
 \begin{equation}
 \phi^{(0)} = \phi _0+ \phi _1z\ .\label{classdilaton}
\end{equation}
This especially simple linear dependence of the dilaton 
is common in papers \cite{{Barrow},{Kanti},{Reb}}. It also 
naturally emerges once one admits the dilatonic dependence only 
on the coordinate along the rotation axis. Since the 
dilaton blows up at the $z$-infinities, they can be condidered 
as additional sources of scalar charge. 

The solution of the Einstein equations can be 
written in the form 
\begin{equation}\label{classolution}
\begin{array}{rcl}
\d s^2&=&\e^{2\alpha }\left(\d t+f\d \varphi\right)^2-
l^2\d\varphi^2\\
&-&\d z^2 -C^{-2}l^{-2}\left(\d \e^ \alpha 
\right)^2\ , 
\end{array}
\end{equation}
the metric functions $f$ and $l^2$ being given by  
\begin{subequations}\label{classmetric}
\begin{eqnarray}
&\displaystyle f=-\frac \Omega C\e^{-2 \alpha }+F\ ,\\ \label{classf}
&\displaystyle C^2 l^2= \Omega ^2\e^{-2 \alpha }+
\frac14 \phi^2_1\e^{2 \alpha}+ D \alpha +E\ ,
\label{calssl^2}
\end{eqnarray}
\end{subequations}
with $\Omega,\ C,\ D,\ E,\ F$ integration constants. 
The physical quantities, the energy density and 
the pressure, are found to be 
\begin{equation}\label{classdensity}
\begin{array}{ll}
&\displaystyle{16\pi\mu=D\e^{-2 \alpha }- \phi^2_1\ ,}\\
&\displaystyle{16\pi p=D	\e^{-2 \alpha }+ \phi^2_1 \ .}
\end{array}
\end{equation}
The formulae \eref{classmetric} and \eref{classdensity} 
are expressed in terms of an arbitrary non-constant $C^2$
function $ \alpha $ that reflects the radial coordinate rescaling 
possibility.  

\section{\label{sec4}First-order solution}

An electromagnetic field is represented by a 2-form $F$ in the 
action \eref{stringaction}. The electromagnetic field, being 
already of $ \varepsilon$-order, is non-minimally coupled to gravity 
with the firm (exponential) dependence on the longitudinal 
direction. Our next task is to take a suitable {\it Ansatz} 
for the electromagnetic field and then solve the equations 
of motion. 
 
Let charge be distributed with a current density $j(r,z)$ 
through a spacetime. Note we have not included the source 
term $A\w*j$, where $A$ is a vector potential, into the action 
\eref{stringaction} because of technical simplicity. This 
is possible if and only if $A\w*j$ is an exact form 
and can be transformed away. 

Of great 
physical importance, in particular on the field of rotating 
spacetimes we deal with, is the case when 
the Lorentz force, in the comoving system 
proportional to $*(u\w*F)$, acting on the fluid 
particles, vanishes. In the fluid rest frame it means 
that only a magnetic field survives. The form 
of the metric field equations of motion, namely 
the $\varphi z$ and $\varphi r$ components, 
leads us to exclude the spacetime with electric currents 
parallel to the axis of rotation, in which the angular 
part of the magnetic field vanishes identically. 
The electromagnetic field 2-form is then given by 
\begin{equation}
F={\rm B}_{\hat r}\ \T 1\w\T 2+{\rm B}_{\hat z}\ \T 3\w\T 1\ .
\label{2-form}
\end{equation}
The presence of the radial magnetic field may seem to be 
artificial because it causes a strange phenomena - an occurence of 
magnetic charges (monopoles). In fact, this is the case. But the form 
of $zr$-component of the Einstein equations, namely the 
equation \eref{zr}, enforces the existence of the radially 
pointing magnetic field in order for the dilaton to be also 
radially dependent. Otherwise it would simply be given by 
\eref{classdilaton}.

The metric field equations of motion following from the 
action \eref{stringaction} are the 
Einstein equations \eref{Einsteinframe} with 
the stress-energy tensor \eref{stress-energy} enriched by the 
electromagnetic field contribution ~\cite{Straumann}, 
where electromagnetic field is non-minimally coupled 
to gravity,  
\[T_{\rm elmag}=\varepsilon\frac{\e^{\phi^{(0)}}}{8\pi}
\ \T \alpha\otimes*
(\F\w i_{\hat\alpha}*\F-i_{\hat\alpha}\ \F\w*\F)\ .\]

The appropriate Einstein-Maxwell system reads
\begin{subequations}\label{Einsteineqs}
\begin{eqnarray} 
&\displaystyle{\frac{\d}{\d r}\left[\frac {\e^{2\alpha}}M
\der f\right]=0\ ,}
  \label{omega}\\[2mm]
&\displaystyle{\!\frac 2M\frac{\d}{\d r}\left[\frac 1M\der \alpha \right]
  \!+\!\frac 1{M^2}\!\left(\pder{\phi}{r}\right)^{\!\!2}\!\!
=\frac{4 \varepsilon}{l^4}\e^{\phi+2\alpha} \Fp 2\ ,}
  \label{ff-rr} \\[2mm]
&\displaystyle{\frac 1M\frac{\d}{\d r}\left[
\frac {\e^{-2\alpha }}M \frac{\d }{\d r}
\left(l^2\e^{2\alpha }\right)\right]=
  32\pi p\, \e^{2\alpha}\ ,}
  \label{zz+rr}\\[2mm]
&\displaystyle
{l^2\pder\phi r\pder\phi z=4 \varepsilon 
\e^ {\phi}F_{r\varphi} \Fp{}\ ,}
  \label{zr}\\[2mm]
&\displaystyle{\frac{\e^{-2\alpha}}M{\frac{\d}{\d  r}
\!\left[\frac1M
  \frac{\d l^2}{\d r}\right]=}
  {16\pi(p-\mu)
  +\frac 1{M^2}\left(\der f\right)^{\!\!2}}\!
-\left(\pder {\phi}{z}\right)^{\!\!2}\!
-4 \varepsilon\e^ \phi \frac{\Fp 2}{l^2}  ,}
  \label{tt-rr} \\
&\displaystyle{2\e^{-2\alpha}\der\alpha 
\frac{\d \, l^2}{\d r}=
32\pi M^2 p- 
  \left(\der f\right)^{\!\!2}+4\varepsilon\e^ {\phi-2\alpha}
F^2_{r\varphi}
-4 \varepsilon \e^ \phi \frac{M^2}{l^2}\Fp 2}\nonumber\\
&\displaystyle{+\ l^2
\left(\pder {\phi}r\right)^{\!\!2}\e^{-2\alpha}
  -\left(\pder {\phi}z\right)^{\!\!2}M^2\ ,}
  \label{intBianchi}
\end{eqnarray}
\end{subequations}
and it is to be completed by the massless Klein-Gordon 
equation \eref{dilaton}, which does not undergo any changes, 
and furthermore by the modified Maxwell equations. 
In \eref{Einsteineqs} as well as in the remainder 
of this section $\e^\phi$ stands 
for $\e^{\phi^{(0)}}$. 

Variation of the action \eref{stringaction} with 
respect to a vector potential $A$ yields the 
generalized Maxwell equations
\begin{equation}\label{Maxwell}
-*\d* (\e^{\phi^{(0)}}\ F)=\frac{4\pi}\varepsilon\ j\ ,
\end{equation}
or in the explicit form
\begin{equation}\label{exMaxwell}
\begin{array}{c}
\displaystyle{l^2\e^{\phi-2\alpha}\pder{}r\left[
(f\delta^\mu_0-\delta^\mu_1)\frac{F_{r\varphi}}M\right]}\\
\displaystyle{+M(f\delta^\mu_0-\delta^\mu_1)\pder{}z(
\e^\phi F_{z\varphi})=-\frac {4\pi}\varepsilon Ml^2j^\mu}\ .
\end{array}
\end{equation}

The same procedure as in the zero-order 
case gives the following system for functions $l^2$ 
and $\alpha$
\begin{subequations}\label{nonlin}
\begin{eqnarray}  
&\displaystyle{2l^4\nder 2{\alpha }m=4\varepsilon
\e^{\phi+2\alpha}F^2_{\varphi z}-{\tilde\phi}^2_2\ ,}
\label{nonlinalpha}\\ 
&\displaystyle{\nder2{l^2}m={\tilde\phi} ^2_1
\e^{2\alpha }+4\Omega^2\e^{-2\alpha }-
4\varepsilon\e^\phi\frac{F^2_{r\varphi}}{M^2} }
\label{nonlinl}
 \end{eqnarray}
\end{subequations}
with new constants ${\tilde\phi}_1$ and 
${\tilde\phi}_2$ which already include the $\varepsilon$-order 
correction.

Because the dilaton is written as \eref{dilcor}
and $\phi^{(0)}$ does not depend on $r$, the term 
$(\pder\phi r)^{2}$ (and also ${\tilde\phi}^2_2$) 
is already of ${\cal O}(\varepsilon^2)$ 
and should be neglected. Putting together equations 
\eref{zr}, \eref{nonlinalpha} 
and \eref{exMaxwell} we have arrived at the 
following conditions for the electromagnetic 
field
\begin{equation}\label{conditions}
\begin{array}{c}
\displaystyle{*(u\w F)\w\T 2=0}\ ,\\
\displaystyle{u\w *F=0}\ .
\end{array}
\end{equation}
The equations \eref{Einsteineqs} and \eref{conditions} 
are solved by a purely longitudinal magnetic field, 
parallel to the rotation axis  
\begin{equation}\label{elmag}
{\rm B}_{\hat z}=B\e^{-\frac12\phi^{(0)}-\alpha}
\ ,\qquad B={\rm const}\ 
\end{equation}
while the radially pointing magnetic field 
vanishes, i.e. ${\rm B}_{\hat r}=0$ in \eref{2-form}.  
As a matter of fact one has quite transparent 
physical interpretation of the found result. 
Since according to \eref{elmag} and \eref{Maxwell} 
it must hold $j\w\T 0=0$, we conclude that the 
fluid particles are the charge carriers, i.e. 
the current density is purely convectional, 
$j=\rho u$. The charge density $\rho$ is 
determined by the formula
\begin{equation}\label{chargedensity}
 4\pi\rho=-\varepsilon \frac BM\der f 
\e^{\frac12\phi^{(0)}-\alpha} \  .
\end{equation}
But there is a price we must pay for the 
simplification. Note that the exterior derivative 
of \eref{elmag} 
does not vanish which means 
that we deal with a current of the 
magnetic monopoles and one has to introduce 
a magnetic charges current density 1-form $ j_m$ by
\begin{equation}
\frac{4\pi}\varepsilon\ j_m=*\d F=-\frac12B\phi_1
\e^{-\frac12\phi^{(0)}-\alpha}\T 0\ .\label{magnetic}
\end{equation}
Equations \eref{chargedensity} and \eref{magnetic} 
show us that the source term $A\w*j$ is identically 
vanishing. 
 
Essentially there are two possibilities to keep 
this situation physically acceptable. Either we can 
expect that going to a non-abelian gauge fields will 
smooth out this solution, or, in the case of abelian 
gauge fields, it is possible to 
introduce $j_m$ explicitly in the action 
\eref{stringaction}, but one has to break the general 
covariance to do this \cite{Schwarz}.

Now we can straightforwardly solve the 
Einstein equations. From the same 
reason as in the zero-order case one 
has one degree of freedom corresponding to the radial 
coordinate rescalling possibility. One finds that one 
has seven
independent equations for exactly eight unknowns: 
four metric functions $f,\ l,\ \alpha ,\ \delta$ and 
four physical quantities $p,\ \phi,\ \mu,\ \rho$. 

The dilaton according to the \eref{dilaton} 
becomes equal to 
\[\phi=\phi_0+{\tilde\phi}_1z\ .\]

The Bianchi identity in our case, provided 
that the scalar field equation of 
motion \eref{dilaton} is fulfilled, is
\begin{equation}\label{Bianchi}
\u\w*(\mu\ \d u+\d(p\ \u)+\varepsilon\ \rho\ {\e^{\phi^{(0)}}}\ \F)=0\ .
\end{equation}

After all one obtains the result \eref{classolution} 
with the following functions $f$ and $l$
\begin{equation}\label{stringl2}
\begin{array}{c}
\displaystyle f=-\frac \Omega C\e^{-2 \alpha }+F\ ,\\ 
\displaystyle C^2 l^2= \Omega ^2\e^{-2 \alpha }+
\frac14 {\tilde\phi}_1^2\e^{2 \alpha}-2 \varepsilon B^2 
\alpha^2+D \alpha +E.
\end{array}
\end{equation}

For the energy density, the pressure and 
the charge density \eref{chargedensity} one has
\begin{equation}\label{stringdensity}
\begin{array}{c}
\displaystyle{16\pi\mu=
\left[D+2 \varepsilon B^2\left(
1-2 \alpha\right)\right]\e^{-2 \alpha }-{\tilde \phi}^2_1\ ,}\\[2mm]
\displaystyle{16\pi p=\left[D-2 \varepsilon B^2\left(
1+2 \alpha\right)\right]\e^{-2 \alpha }+ {\tilde\phi}^2_1 \ ,}\\[2mm]
\displaystyle{2\pi \rho=-\varepsilon \Omega B
\e^{\frac12\phi^{(0)}-3\alpha}\ .}
\end{array}
\end{equation}

The mathematical structure of the solution \eref{stringl2} 
and \eref{stringdensity} of the 
Einstein-Maxwell equations \eref{Einsteineqs}
is much the same as the zeroth-order one \eref
{classmetric} and \eref{classdensity}. 
The differences 
appear in the presence of the term 
quadratic in $\alpha$ in the function $l^2$ 
in \eref{stringl2} and the linear terms in 
$ \alpha $ in \eref{stringdensity}, which 
occurs due to the existence of the 
magnetic field. The function $f$ remains 
unchanged. 

\section{\label{sec5}Case with more scalar fields}

It is straightforward to generalize 
the zeroth order solution \eref{classmetric} 
and the first order solution \eref{stringl2} 
in the case when more scalar fields are present. 
The motivation comes from string theory, where 
it is known that the effective 
description at low energies may contain not only 
the dilaton, but also others tensor fields, depending 
on how the compactification was carried out \cite{Witten}. 
Among them is most important an axionic tensor field that 
can be, just in four dimensions, represented by 
an extra massless scalar field. Also some 
additional massless scalar fields, 
called modulus fields, may be present \cite{Kanti}.   

We shall consider $N$ massless scalar fields $\phi_i$, 
$i=1,2,..N$, and $N$ non-minimally coupled massless scalar fields 
$\psi_i$. 
The total action 
\eref{stringaction} can be rewritten as 
\begin{eqnarray}
S_{\rm eff}&=&\intop_{\cal M}\left[*R+16\pi*\mu
+2\varepsilon\ \e^{\phi}\ F\wedge *F\right.\nonumber\\
&-& \sum_i(\d \phi_i\w *\d \phi_i+
\e^{-2\phi_i} \d \psi_i\w *\d \psi_i)] \ .
\label{stringactiongen} \end{eqnarray}

\noindent Before proceeding further let us mention that 
each scalar field $\phi_i$ or $\psi_i$ can be 
written a similar way to equation \eref{dilcor}. 

\subsection{Zeroth order in $\varepsilon$}

The Klein-Gordon equation \eref{Klein-Gordon} 
for each of the scalar fields $\phi_i$ still holds, 
while the equations of motion for the scalar fields $\psi_i$ 
are given by 
\begin{equation}\label{Klein-Gordon,axion}
*\d (\e^{-2\phi_i}*\d \psi_i)=0\ ,
\end{equation}
for each index $i$. 
The modified Einstein's field equation are listed 
below. Again, as in section \ref{sec3}, the authors 
were able to solve the generalization of 
\eref{nonlinclass} provided that neither $\phi_i$ 
nor $\psi_i$ depends on the radial coordinate. 
Then from \eref{tt-rr,gen} we have 
\begin{equation}
\phi_i=\phi_{i0}+\phi_{i1}z\ ,\ 
\psi_i=\psi_{i0}+\e^{\phi_i}\psi_{i1}\ ,\label{dilax}
\end{equation}
which inserted into the \eref{Klein-Gordon,axion} 
yields $\phi_{i1}\psi_{i1}=0$. In \eref{dilax} 
$\phi_{i0},\phi_{i1},\psi_{i0},\psi_{i1}$ 
are integration constants. Thus the only non-trivial 
zero-order solution is given by 
\begin{equation}\label{dilclass,gen}
\phi^{(0)}_i=\phi_{i0}+\phi_{i1}z\ ,\ 
\psi^{(0)}_i=\psi_{i0}\ .
\end{equation}
The solution \eref{classmetric} and \eref{classdensity} 
remains unaffected provided $\phi^2_1$ is replaced by 
$\Phi^2_1=\sum\phi^2_{i1}$.  

\subsection{First order in $\varepsilon$}

As a consequence of the presence of more 
massless scalar fields, one has to modify the 
equation \eref{Einsteineqs} in the following manner. 
The equation \eref{ff-rr} becomes 
\begin{eqnarray}
\frac 2M\frac{\d}{\d r}\left[\frac 1M\der \alpha \right]
-\frac{4 \varepsilon}{l^4}\e^{\phi+2\alpha} \Fp 2\nonumber\\
 = -\frac 1{M^2}\sum_i\!\left[\left(\pder{\phi_i}{r}\right)^{\!\!2}
+\e^{-2\phi_i}\left(\pder{\psi_i}r\right)^{\!\!2}\right]
\ .\label{ff-rr,gen}
\end{eqnarray}
The term $(\pder \phi z)^{2}$ on the right-hand side 
of the equation \eref{tt-rr} should be replaced by 
\begin{equation}\label{tt-rr,gen}
\sum_i\left[\left(\pder{\phi_i}{z}\right)^{\!\!2}
+\e^{-2\phi_i}\left(\pder{\psi_i}z\right)^{\!\!2}\right]\ .
\end{equation}
Similarly the equation \eref{intBianchi} will be changed 
in an obvious way. Finally \eref{zr} becomes 
\begin{equation}\label{zr,gen}
l^2\sum_i\!\left(\pder{\phi_i}{r}\pder{\phi_i}z+
\e^{-2\phi_i}\pder{\psi_i}r\pder{\psi_i}z\right)=
4\varepsilon\e^{\phi}F_{r\varphi}F_{\varphi z}\ .
\end{equation}

It was stated before that the scalar fields are decomposed 
into zero-order part and first-order correction as 
\[
\phi_i=\phi_{i0}+\phi_{i1}z+\varepsilon\phi^{(1)}_i \ ,\ 
\psi_i=\psi_{i0}+\varepsilon\psi^{(1)}_i \ .\]
As a matter of fact it is seen that all terms in the 
modified Einstein's equations involving 
$(\pder{\psi_i}r)^{2},(\pder{\psi_i}z)^{2}$ 
and even $(\pder{\phi_i}r)^{2}$ are already of 
${\cal O}(\varepsilon^{2})$ and have to be ignored. Therefore 
we continue to keep our {\it Ansatz} \eref{elmag} 
for the electromagnetic field. From this fact 
it immediately follows that the term $F\w F$ 
vanishes identically. The equation 
\eref{magnetic} should be modified due to the 
presence of the zero-order axionic field. 
But $\psi^{(0)}=\psi_0$ is constant, which 
can be set equal to one, and so in particular 
the equation \eref{magnetic} applies in this case as well. 
It also means that, 
for example, from \eref{zr}, the 
scalar fields $\phi_i$ are given by
\begin{equation}\label{zbytec}
\phi_i=\phi_{i0}+{\tilde\phi}_{i1}z\ ,
\end{equation}  
with constants ${\tilde\phi}_{i1}$ including 
the $\varepsilon$-order corrections to $\phi_{i1}$. 
The resulting metric and physical quantities 
are still given by \eref{stringl2} and \eref{stringdensity} 
provided $\ph 2$ is replaced by 
${\tilde\Phi}^2_1=\sum{\tilde\phi}^2_{i1}$. 

Since the Einstein's equations give us no information 
with respect to the fields $\psi_i$ one has to 
solve their equations of motion \eref{Klein-Gordon,axion}. 
One can carry out the separation of variables to obtain 
\begin{equation}\label{axion}
\psi_i=\psi_{i0}+\varepsilon\e^{\phi_{i1}z}
[A_i\cos(v_iz)+B_i\sin(v_iz)]\ \eta_i(r)\ ,
\end{equation}
where $A_i,\ B_i$ and $v_i$ are arbitrary constants 
and the functions $\eta_i$ are solutions of 
second-order equations that can be transformed into 
the form
\begin{equation}\label{radialaxion}
\e^{-2\alpha}\derm{}\left(l^2\derm{\eta_i}\right)
-(\phi_{i1}^2+v_i^2)\ \eta_i=0\ .
\end{equation}

\section{\label{sec8}String-inspired theory of gravity}

The aim of this section is to show that the 
solution described by the metric \eref{stringl2} 
actually remains unaltered even if string-inspired 
charged gravity is taken under consideration 
\cite{Kanti}. 
 
The total string-inspired effective action 
\eref{stringaction} can be rewritten as \cite{{Kanti},{Rizos}}
\begin{eqnarray}
S_{\rm eff}&=&\intop_{\cal M}\left[*R+16\pi*\mu\right.\nonumber\\
&-&\left. \sum_i(\d \phi_i\w *\d \phi_i+
\e^{-2\phi_i} \d \psi_i\w *\d \psi_i)\right.\nonumber\\
&-&\left. 8\pi^2\varepsilon \e^\phi e({\cal M})
+4\varepsilon\ \psi\ {\rm Tr}\ \Omega\w\Omega\right. \nonumber\\
&+&\left. 2\varepsilon\ \e^{\phi}\ F\wedge *F-4\varepsilon
\ \psi\ F\w F\right]\ .\label{stringactiongen2} 
\end{eqnarray}
In terms of the 
inverse string tension $\alpha'$ (Regge 
slope) and the string coupling constant 
$\rm g$ the parameter $\varepsilon$ is expressed like 
$\varepsilon=\frac{\alpha^\prime}{4{\rm g}^2}$.  
The Euler class 
$e(\cal M)$ of $T{\cal M}$ occuring in 
\eref{stringactiongen2} is in four dimensions 
equal to \cite{Witten}
\begin{equation}\label{Euler}
e({\cal M})=\frac 1{8\pi^2}
\left(R_{\alpha\beta\gamma\delta}R^{\alpha\beta\gamma\delta}-
4R_{\alpha\beta}R^{\alpha\beta}+R^2\right){\eta}\ ,
\end{equation}
$\eta$ is the volume element with components $\eta_{
\alpha\beta\gamma\delta}$. 

We have also added 
an extra contribution arising from a field $\mu$. 
Physically it may represent an energy density of a fermionic 
matter, that is in our model approximated by a perfect fluid 
of a pressure $p$. Although this picture is rather intuitive 
and is not as transparent as in former EMD, 
later it will turn out to be useful. 
If one wants to supress the fermionic matter and 
recover an ordinary string-inspired action with 
the cosmological constant 
$ \Lambda$, then our approach is fruitful too since the 
state equation $\mu+p=0$ along with $ \Lambda=-8\pi p$ 
gives the desired modification of the action.  
The fields $\phi_N\equiv \phi$ and $\psi_N\equiv
\psi$ may be referred to as the dilaton and axion respectively. 

For any cylindrically symmetric stationary metric, i.e. 
metric depending on the radial coordinate with the only 
non-vanishing cross-term 
$g_{\varphi t}$, a straightforward calculation gives the 
following useful formula for the Euler class \eref{Euler} 
\begin{equation}\label{GB} 
\begin{array}{rcl}
e({\cal M})\!&=\!&\displaystyle{\frac1{4\pi^2}\frac{\d }{\d r}}
\left\{\left[ \left(\der f\right)^2e^{2\alpha}\!
+2\der\alpha\der {l^2}\right]
l^{-1}\e^{\alpha+\gamma-3\delta}\der\gamma\right\}\\[2mm] 
&\times& l^{-1}\e^{-\alpha-\gamma-\delta}\ 
\T 0\w\T 1\w\T 2\w\T 3\ .
\end{array}
\end{equation}
In \eref{GB} the function $\gamma$ is given by 
$g_{zz}=-\e^{2\gamma}$. Thus for 
\eref{classolution} the Euler class vanishes. 
For the basis \eref{basis} it turns 
out that the term 
\[ 
{\rm Tr}\ \Omega\w\Omega=-\frac14R^{\alpha\beta}_{~~~\gamma\delta}
R^{\sigma\tau}_{~~~\alpha\beta}\ \eta^{\gamma\delta}_
{~~~\sigma\tau}\ \eta\]
is identically vanishing.

Since the magnetic field \eref{elmag} 
is purely longitudinal, the term 
$F\w F$ is vanishing too and our 
problem in fact reduces to the Einstein-Maxwell-dilaton plus 
fluid system disscussed in previous sections. 
This completes the proof that the metric \eref{stringl2} 
after appropriate replacement
$\ph 2\rightarrow{\tilde\Phi}^2_1=\sum{\tilde\phi}^2_{i1}$ 
constitutes string-inspired solution. The scalar fields are given 
by \eref{zbytec} and \eref{axion}, subject to the 
equation \eref{radialaxion}.

\section{\label{sec6}On some atributes of the solution}

We briefly comment on some physical atributes of the solution 
\eref{stringl2}. 

Of course no every specialization of the 
integration constants in \eref{stringl2} 
leads to cylindrically symmetric spacetime. 
If one requires the found solutions to be 
cylindrically symmetric and regular at the origin 
the axial symmetry condition and the elementary 
flatness condition have to be imposed \cite{Kramer}.
In our case, provided $\alpha\propto r^2$ for small 
vaues of $r$, these conditions give
\begin{equation}\label{regularity}
\begin{array}{c}
\displaystyle{\Omega^2-\frac14{\tilde\phi}_1^2-\frac D2=\pm\ C}\ ,\\[1mm]
\displaystyle{\Omega^2+\frac14{\tilde\phi}_1^2+E=0}\ ,\\[1mm]
\displaystyle{F=\frac \Omega C}\ .
\end{array}
\end{equation}

Clearly the Lorentz force is vanishing. Also it can 
straightforwardly seen that the source term $A\w*j$ is 
exact form. Furthermore non-geodesic motion of the fluid 
should be understood as a mere consequence of a pressure 
inhomogeneity. Indeed it follows from fact that the 
accelaration is ${\dot u}=-\d\alpha$ and that the Bianchi 
identity, \eref{Bianchicl} or \eref{Bianchi}, can be rewritten as
\[\left(\mu+p\right)\der\alpha=-\der p\ .\]

\noindent The vorticity 1-form $\omega$ according to 
\eref{vorticity} equals 
\[{\omega}=\frac12*(\u\w\d \u)=\Omega\e^{-2\alpha}\d z\ .\] 
In this way the metric \eref{stringl2} is static if 
and only if $\Omega=0$. 

Let us also write down how the energy 
conditions restrict the ranges of the 
integration constants in \eref{stringl2}. 
The strong along with the dominant energy 
condition imply the following two inequalities
\begin{equation}\label{EC}
\begin{array}{c}
\displaystyle{D-4\varepsilon B^2\alpha+\frac12\ph 2\e^{2\alpha}\geq0}\ ,\\[3mm]
\displaystyle{4\varepsilon B^2\geq\ph2\e^{2\alpha}}\ .
\end{array}
\end{equation}
 
Next remark concernes the algebraic 
classification of the Weyl tensor. It turns 
out that the metric \eref{stringl2} is of 
Petrov type $D$ except on the hypersurfaces 
(one or more) given by 
\[4\varepsilon B^2\alpha=2\varepsilon B^2+D\ ,\]
where it is of type $O$. 

Last remark clarifies the connection between 
\eref{stringl2} and G\"odel-type solutions \cite{Reb}, 
provided that the function $ \alpha$ and 
integration constants in \eref{stringl2} are 
chosen conviniently. 
The following appropriate choice respecting 
regularity conditions \eref{regularity} has been done 
\begin{equation}\label{choice}
\begin{array}{l}
\displaystyle{\Omega^2=\varepsilon B^2+\frac1{2a^2}
-\frac14\ph2\ ,\ E=-\left(\frac1{2a^2}+
\varepsilon B^2\right)}\ , \\
\displaystyle{D=\frac1{a^2}+2\varepsilon B^2-\ph2+2C\ ,
\ F=\frac \Omega C}\ .
\end{array}
\end{equation}
The physical meaning of the constant $a$ will be 
clear shortly. 
Let us now specify the arbitrary function 
$\alpha$ as $\alpha=2a^2C{\rm sh}^2(\frac r{2a})$. 
For simplicity let us consider only the dilatonic and 
axionic fields. 
We obtain new metric that depends upon three parameters: 
$a, \ B$ and $C$ (explicit form is 
omitted here). This solution describes an inhomogeneous 
universe and from \eref{EC} it is immediate that $C$ must 
not be positive. If $C$ is set equal to zero, 
the general solution \eref{stringl2} becomes 
\begin{equation}\label{Kanti}
\!\d s^2\!=\![\!\d t\!+\! 4a^2\Omega 
{\rm sh}^2(\frac r{2a})\d \varphi]^{2}\!-\!
a^2{\rm sh}^2(\frac ra)\d \varphi^2
-\d z^2-\d r^2,
\end{equation}
which is manifestly of the G\"odel-type. If in addition 
the state equation $p+\mu=0$ is requested, the scalar field 
contribution must have the form 
\begin{equation}\label{Kantidil}
\ph 2= \frac1{a^2}+2\varepsilon B^2\ .
\end{equation}
The latter 
equation reflects fact that the non-vanishing cosmological 
constant rather then the perfect fluid is considered. If 
we let $\varepsilon\rightarrow 0$ then from the section 
\ref{sec3}, especially from the equation \eref{classdilaton}, 
one has $a^2\phi_1^2=1$, and from \eref{dilcor}, \eref{zbytec} 
and \eref{Kantidil} the $\varepsilon$-order corrected 
dilaton is given by    
\begin{equation}\label{Kantinormal}
\phi=\phi_0+\phi_1z\left(1+\varepsilon B^2a\right)\ .
\end{equation}
The relationship between fundamental parameters of the 
theory becomes 
\begin{equation}\label{fund}
4\Omega^2-\frac1{a^2}=2\varepsilon B^2\ ,
\end{equation}
subject to the inequality $2\varepsilon a^2B^2\geq 1$, 
which follows from the energy conditions \eref{EC}. 
 
For the axion one has the equations \eref{axion} and 
\eref{radialaxion}. With help of the elementary theory 
of Legendre polynomials we find $v=\phi_1$ and
\[\displaystyle{\psi=\psi_0+\varepsilon\e^{\phi_1z}{\rm ch}
\left(\frac ra\right)
\left[A\cos\left(\frac za\right)
+B\sin\left(\frac za\right)\right]}\ .\]

Note that in zero-order regime $\varepsilon\rightarrow 0$ 
one has $a^2\phi^2_1=1$ and $4\Omega^2=a^{-2}$, 
the latter equality immediately implying 
$g_{\varphi\varphi}\leq 0$. Thus there are no 
CTC's in the spacetime. On the other hand in 
$\varepsilon$-order framework, since 
because of \eref{EC} is $\varepsilon$ generally non-negative, 
$g_{\varphi\varphi}$ 
becomes positive for sufficiently large $r$. 
Thus the first 
order corrections cause the chronology violation. 

Equations \eref{Kanti}, \eref{Kantinormal} and \eref{fund} 
are identical with these of Kanti and Vayonakis 
\cite{Kanti} in string-inspired charged gravity 
framework, when $\varepsilon=\frac{ \alpha'}{4{\rm g}^2}$. 
Their $\alpha'$-order solution arising from Som and Raychaudhuri 
{\it Ansatz} for the electromagnetic field 
turned out to be most favored between 
others cases, also belonging to the $\alpha'$-order. 

We could also obtain proper generalization 
of another solution in \cite{Kanti} with 
a positive cosmological constant, simply by 
carrying out the imaginary transformation 
$a\rightarrow ia$ in \eref{choice}, but we will 
not follow this line further.
 
\section{\label{sec7}Discussion}
In the paper a class of stationary symmetric 
spacetimes within the framework of 
Einstein-Maxwell-dilaton gravity was found, 
that exhibits cylindrical symmetry. 
This solution is exact up to first order 
in parameter $\varepsilon$.  
Provided that the scalar fields in zeroth order 
do not depend on the radial coordinate we 
were also able to find the generalization to 
the case when more scalar fields are present.

The solution obtained depends upon the dilaton 
gradient and the magnetic field in the 
longitudinal direction. It is worthwhile to note that 
the Gauss-Bonnet term vanishes. 
Namely for this reason is found metric exact 
$\alpha'$-order solution in string-inspired theory 
framework. 

Since the forms of zero and first order solutions are 
similar, we can straightforwardly find out what is 
a consequence of the electromagnetic field 
presence with respect to the chronology violation. 
It turns out that it has the power to break down the 
chronology even if the zero-order solution was 
chronologically well behaved. 

\section*{Acknowledgments}
The authors are obligated to Dr Rikard von Unge for 
many helpful and enlightening disscussions. This work 
was supported by grant 201/00/0724. PK also acknowledges to 
Prof. Graham S Hall and to NATO grant RGF0042.

\end{document}